# Correcting self-intersecting polygons using minimal memory
## A simple and efficient algorithm
## and thoughts on line-segment intersection algorithms

**Jean Souviron[1]**

COGITECH Jean Souviron
613 D'Ailleboust
Montréal, Quebec,
H2R 1K2 Canada

*JeanSouviron@hotmail.com*

**Abstract:** While well-known methods to list the intersections of either a list of segments or a complex polygon aim at achieving optimal time-complexity they often do so at the cost of memory comsumption and complex code. Real-life software optimisation however lies in optimising at the same time speed and memory usage as well as keeping code simple. This paper first presents some thoughts on the available algorithms in terms of memory usage leading to a very simple scan-line-based algorithm aiming at answering that challenge. Although sub-optimal in terms of speed it is optimal if both speed and memory space are taken together and is very easy to implement. For $N$ segments and $k$ intersections it uses only $N$ additional integers and lists the intersections in O($N^{1.26}$) or corrects them in O($(N+k)\ N^{0.26}$) at most in average, with a high probability of a much lower exponent around 0.16 and even as low as 0.1. It is therefore well adapted for inclusion in larger software and seems like a good compromise. Worst-case is in O($N^2$). Then the paper will focus on differences between available methods and the brute-force algorithm and a solution is proposed. Although sub-optimal its applications could mainly be to answer in a fast way a number of scattered unrelated intersection queries using minimal complexity and additional resources.

**Keywords:** complex polygon, simple polygon, self-intersecting polygon, optimisation methods, line-segment intersection, sweep-line, scan-line.

---

[1] Dr Jean Souviron, Ph.D.1984, has been an independent consultant in scientific programming since 1994..

## 1. Introduction

Extensive work has been done over the years on the subject of detecting line-segment intersections. As some of the stepping stones in this field one can cite both Shamos & Hoey[8] in 1976 and Bentley & Ottman[3] in 1979. Both algorithms were based on a sweep-line method, i.e. moving a line along an ordered list of segments extremities. Then several new schemes were derived, most if not all of them also sweep-line based, like the famous Chazelle & Edelsbrunner[4] in 1982, and more recently Balaban[2] in 1995, Chen & Chan[5] in 2003 or Eppstein & al.[7] in 2009.

All these methods have been focusing on reaching optimal time complexity. As a means to this goal most used binary tree structures of some sort as a starting point, whether it be balanced (*e.g. Chen and Chan*), red-black trees (*e.g. Chazelle & Edelsbunner*) or some other form, while Eppstein & al. use a Voronoi diagram. Then some have used priority queues (*e.g. Bentley & Ottman and all others deriving from their work*). However even though in terms of space complexity some methods are in O($N$) the Big-O notation hides the constant factor, which nevertherless induces a sometimes not negligible overhead. Finally a few methods claiming to be in O($1$) space complexity are fairly difficult to implement (*e.g. Chen & Chan*).

While Shamos & Hoey looked for a test of whether a polygon was simple or self-intersecting, most if not all of these methods were directed at *listing* the intersections of a set of disjoint segments (*e.g. Bentley & Ottman, Chazelle or Balaban*) while some aimed at polygon decomposition (*e.g. Eppstein & al. or Arkin & al.[11]*). Although applying these algorithms to *correct* a self-intersecting polygon should be expected to be relatively easy, whether through iterations or some additional computations and/or backtracking,  the queue deletion process (*for the queue-based algorithms*) as well as some implementations referring to the segments's numbers will have to be modified to take into account the implied re-ordering or re-numbering of vertices. Some of these methods however propose to correct the self-intersections by simply adding two points at the intersection, like what tools like the ArcGIS *Repair* or the OpenGL *tessellator* do. This is however a pure geometric reasoning making sense only in order to *draw* such a polygon but it will lead to misuses in a general approach where the intersection point has no meaning in itself, like what happens if it orignates from a computational side-effect, for it creates two (*or more*) separate polygons from a single one.

Finally, although solving the problem at hand, these methods do not give the same end-result than the brute-force algorithm, a fact somewhat unusual in computational geometry.

## 2.  Memory usage

### 2.1 General analysis

Different data structures are used for the building and update of the trees involved in all these methods. Some only store node indexes, number of children, and left/right properties while others store more parameters in order to reduce the number of later computations, such as the slope of the segment. *At minima* therefore for $N$ points they use *3N* integers but usually much more. Some methods like Chen & Chan's have a complex encoding scheme to reduce the space needed to store the information. Then algorithms derived from Bentley & Ottman use at least one priority queue containing at the minimum 2 integers per extremity, thus *2N* integers more. So *at minima* these algorithms need *5N* (*sometimes up to 8 or 10 N*) integers to process $N$ points, not taking into account the array needed to sort the data.

       JeanSouviron@hotmail.com

However if this algorithm is to be included in a larger and more complex one this might prove quite a burden on the overall performances. These methods also need a fair amount of additional code, some of it simple, as the building and handling of trees, while some other parts are quite complex like what is described in Chazelle & Edelsbrunner. Finally although the time complexity optimisation is always essential, in times where multithreading, concurrent processing, embedded software and gigantic net-based or net-related databases are increasingly part of the computational environment, devoting such an amount of memory to obtain optimal speed might not be the only factor to take into account.

Therefore the objective to obtain a maximum speed optimisation while using minimal memory space and avoiding too much additional code might prove to be worthwhile. The author limited the study to polygons. In one way they are simpler to handle than disjoint segments as each beginning of a segment is the end of the previous one. On the other hand they are more complex, as their vertices are ordered and thus reversing the order has impacts beyond the two segments involved in the intersection.

For all above-mentioned methods the starting point is to sort the segments's extremities by increasing value of a coordinate. This will lead to the well-known Figure 1.

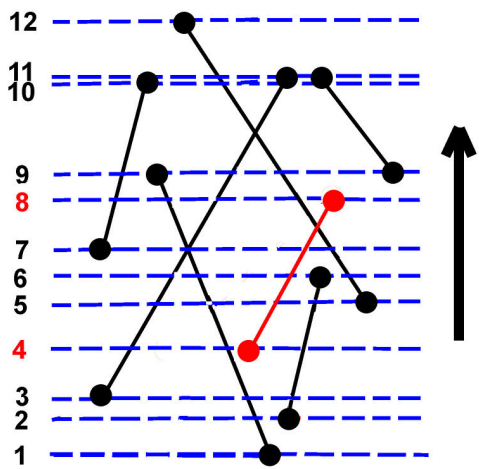

**Figure 1.** Sorted segments's extremities

*The vertical arrow indicates the sorting direction.*

The idea is to use a sweep-line, i.e. a virtual line going through the points, keeping some useful information as to reduce the number of computations and predict whether two segments *could* intersect as the line moves. In order to optimise the speed of search, insertion and deletion processes, trees are used to store the initial points and possible candidates.

However the idea of a sweeping-line going at the same time up and from left to right originates, as was mentioned by Shamos & Hoey, from the less sophisticated scan-line approach used during the previous years during which memory was scarce.

Assuming that the algorithm processes the sorted input sequentially, once the position 4 in Figure 1 is reached, then all potential candidates for an unsolved intersection with the segment 4-8 would lie *in between* the segment's extremities: a segment whose lower extremity is lower would have been already checked (*e.g segment 2-6 or 3-11*), and a segment whose lower extremity is higher will not have any possible intersection (*e.g. segment 9-11*).

Thus if the scan-line approach was to be used one would have to check for all segments for which an extremity lies in the interval defined by the segment's extremities. In the above-mentioned case all segments lying in the interval 4-8 are potential candidates. However as segment 2-6's lower extremity is lower than the position 4 checking can be avoided. But the first encountered candidate, in posiiton 5, will define an intersection and the exploration will consequently stop.

Obviously while doing that process for all the points one will check the same point several times: segment ranges are overlapping. This will by definition lead to the fact that the number of explored points per segment will be a fractional-power part of the total number of points. It is in order to avoid that factor that the listed methods use trees and priority queues.

### 2.2 Outline of the basic scan-line algorithm

In a polygon whose vertices were sorted along one direction, a vertex can be the origin of either 0, 1 or 2 segments with higher extremities.

The basic routine is thus to determine for a given vertex the number of adjacent vertices lying above in the sorted array. If not zero a loop going through these segments then explores the potential candidates. The basic routine will then be applied to these candidates and for each found segment a check is made of whether it intersects the studied segment.

The algorithm to *report* all intersections is thus straightforward as pseudo-code below shows (*sorting is not mentioned).*

```
Loop using p from 1 to N

   Nsegs = Finds higher extremities for p

   Loop using q from 1 to Nsegs

       Loop using r from (p+1) until index(q)
          Nsegs1 = Finds higher extremities for r
          Loop using s from 1 to Nsegs1
             If Intersect (seg(r,s), seg(p,q))
                Reports intersection
             Endif
          Endloop
       Endloop

   Endloop

Endloop
```

However if the algorithm were to be used to *correct* intersections not only will it have to stop as soon as one intersection is found, correct it, and backtrack to start again, but subtle differences will have to be introduced to take into account all possibilities of newly created intersections.

Apart from the average case where correcting an intersection will lead to either no new intersection for the studied segment or to a new intersection but with a segment lying above in the sorted array, which will be detected while backtracking normally, three special situations can appear and are detailed below and in Figure 2.

          JeanSouviron@hotmail.com

- First as a segment might have been "shortened" in the sorted array, i.e. the new upper position might be lower than what it was before the correction, when backtracking to the same position the upper limit of range exploration has to be set up to this former position and not to the actual ending position (*Figure 2a & 2b*).

- Then because of the re-ordering of segments when an intersection is corrected, when the algorithm has backtracked to the same position the range exploraton has to be modified to take into account segments for which one extremity lies *below* the lowest position of the actual studied segment (*i.e. allowing backtracking in the sorted array*) (*Figure 2c & 2d*).

- Finally as it is a polygon and the lowest impacted point in the array might be lower than the actual position and as a vertex has two adjacent segments, one of which could be below, the backtracking should not go to the same position but rather to the position of this lowest extremity of the lowest impacted point if it exists (*Figure 2e & 2f*).

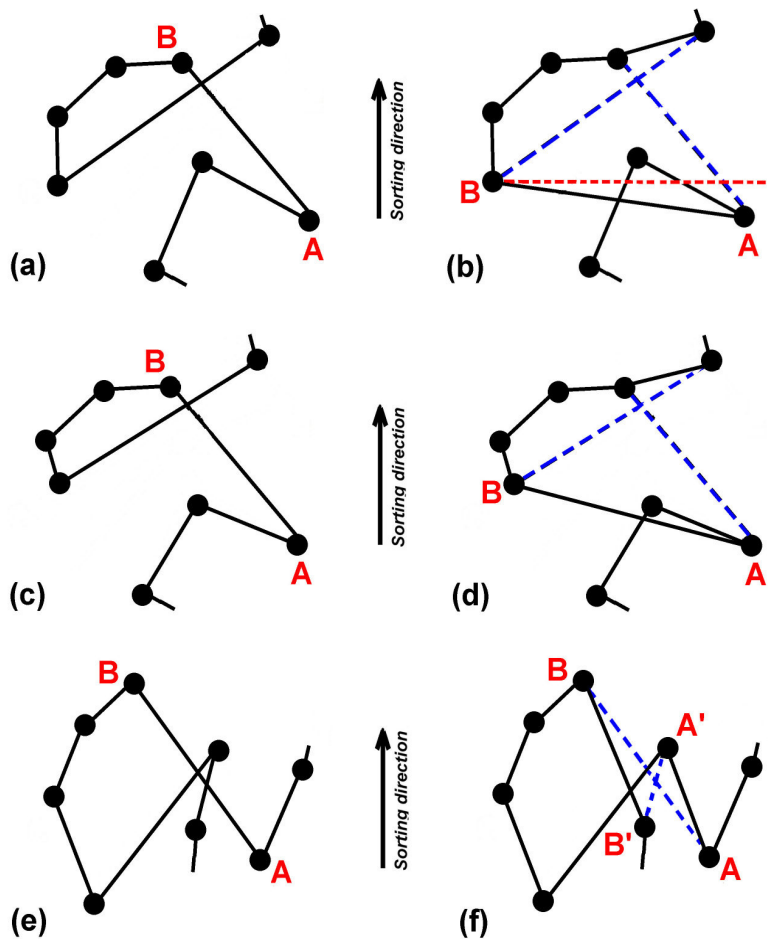

**Figure 2.** The three possible special cases after having resolved an interection in a polygon

*AB represents the studied segment, A being the actual position in the sorted array. Blue dashed lines are for the former segments involved in the intersection.*

So to summarize, if a previous intersection was corrected the algorithm should:

- allow exploration to reach former high end of the segment when the same position is reached.

- allow exploration to backtrack in the sorted array when the position lies in the last inpacted interval

and once an intersection is detected it has to:

- check whether the other extremity of the candidate segment lies lower than the actual position if backtracking was allowed,. If such is the case then it has to find the lowest extremity of its originating segments and backtrack to this position or simply backtrack to the same position otherwise.

- find the highest impacted index in the array.

In consequence the final algorithm to *correct* all intersections is shown as pseudo-code below

```
Loop using p from 1 to N-1

    Nsegs = Finds higher extremities for p

    If p equals the last position where an intersection occurred
        Sets exploration upper limit to former upper limit
    Else
        Sets default exploration (up to normal end of segment)
    EndIf

    If p lies in the interval defined by the last limits
        Allows backtracking
    Else
        Forbids backtracking
    EndIf

    Loop using q from 1 to Nsegs
        Loop using r from (p+1) until upper limit
            Loop using s from 0 to 1
                Finds other extremity of segment r-s
                If backtracking is forbidden and extremity is lower than r
                    Skips
                Else
                    If Intersect (seg(r,s), seg(p,q))
                        Stores intersection
                        Exit
                    EndIf
                EndIf
            Endloop
        Endloop
    Endloop

    If intersection is found
        Stores actual position p
        Stores the position of q

        Corrects intersection

        Sets high interval limit to r
        If position of s < p    (backtracking)
            Sets low interval limit to this index
            Backtracks to position of lowest extremity ending at s
        Else
            Sets low interval limit to p
            Backtracks to actual position
        EndIf
    EndIf

Endloop
```

Please note that sorting is not mentioned in the pseudo-code. Furthermore the true computation of whether two segments intersect can be avoided if their intervals in the direction perpendicular to the sorting's one do not overlap.

Finally a last technical note should be made: if the coordinates are integer values, or represent integer values, it is essential for the backtracking position to be at the starting point of the same-value range in the sorted array, for the order of points implied by the sorting might induce left-over segments if one only uses the segment's extremity index.

### 2.3 Special caution

The algorithm to correct a self-intersection in a polygon is well-known and straightforward: if an intersection is found involving segments ($i$, $i+1$) and ($i+x$, $i+x+1$), the points with indexes in the interval [$i+1$, $i+x$] are to be put in reverse order. However if an intersection involves the second point (*the one indexed "1"*) or the last, it might result in modifying the orientation of the polygon if it involves also almost-symmetrical points, i.e. points lying at the other end of the array. This is not desirable in general. Therefore a test before starting the computations (*e.g. when determining the major coodinate*) and at the end of it must be performed and eventually a reversal of the points between the second and the last one should be done.

Also worth mentioning is the fact that as the reference is the indexes of the points the update of the sorted indexes should be applied before the real update of the points.

### 2.4 Worst case of a scan-line approach

In a scan-line approach worst-case consists in having to explore most of the points for each segment. Such a case is detailed in Figure 3. This corresponds to a case where most segments overlap along the sorting direction.

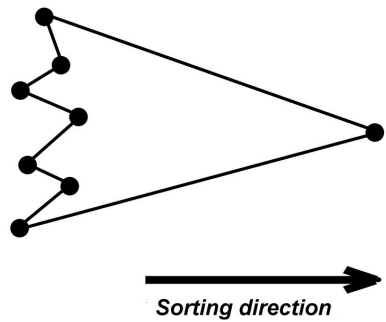

**Figure 3.** Worst case of a scan-line approach

In such a case time complexity is in O($N^2$) as for each segment the algorithm has to check all the remaining segments. It could even be worse than a brute-force algorithm which, although also in O($N^2$), would only need *N/2* comparisons per segment in average.

### 2.5 Average case of a scan-line approach

In order to evaluate the potential of the above-mentioned simple approach the *value* of the fractional-power factor involved is essential. Some effort was spent to obtain test data in large numbers as to eventually derive global experimental figures. First raw datasets were used. They come from a variety of origins and cover a wide range in the number of points and distributions. They are formed from lightning data[1], subsets of public geo-political information files[2], medical images[3], subsets of some botanical data[4], two geographical maps[5] and computer-generated examples of clusters used for research purposes[6]. As a whole they form 790 datasets containing from 4 up to more than 760,000 points. The polygons in the present study were obtained through the most detailed settings of the "*Naked-Eye*" algorithm described in Souviron[9] and contained from 4 up to more than 105,000 vertices. Although lightning datasets are the most random and do not lead to *any* worst-case, in some of the other sets there were some, as Figure 4 shows..

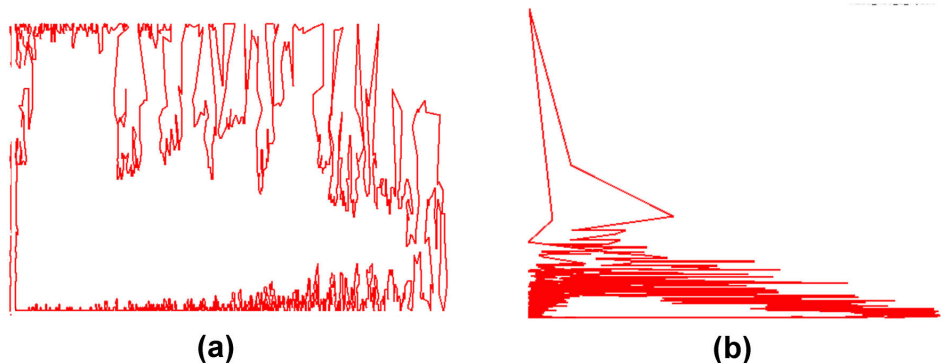

**(a)**                    **(b)**

**Figure 4.** Examples of worst-cases polygons found in some datasets

*(a) botanical  (b) geopolitical*

After the removal of 34 of such worst-cases a total of 756 self-intersecting polygons remained. Although the *number* of self-crossings per polygon is strongly related to the algorithm used to create the polygons, the polygons themselves are a good sample of average cases containing a large variety of polygon shapes.

Then in order to confirm the results and remove even the remote possible influence of the original algorithm by using usual polygon sources a series of completely independent polygons was also used. They are GIS-related polygons[7] and form a sample of 14,425 self-intersecting polygons containing from 4 up to 3297 vertices. In consequence the figures and limits presented in Figure 5 are representative of the average case. It is worth noting from Figure 5b that on the 14,425 GIS-polygons only 3 fall above the high limit obtained from raw data, and even then they are not very far away. They also appear to have an even lower exponent factor (*0.11*). As the algorithm used to build the polygons outputs an extremely noisy contour because of the algorithm's most detailed settings (*see Figure 6*) it thus could be concluded that for the average use the value given by the GIS-related polygons is a realistic one, but nevertheless an upper bound for average situations is of the order of magnitude of the quadratic root of $N$. As such, although not optimal, this method provides a very good speed optimisation while keeping a very low memory usage and a very simple code.

---

[1] Lightning strike locations obtained in two days in the summer of 1998 through the CLDN (*Canadian Lightning Detection Network*), courtesy of Environment Canada. Selected within time bins (*from 10 minutes up to 2 hours*) and resolution bins (*from 2.5 up to 350 km minimum distance between locations*), they form a sample of 629 datasets, ranging from 4 to more than 93,000 points.

[2] RGC dataset (*France's cities geographic directory*) of IGN (*french National Geographic Institute*). 31 files were obtained by selecting several population ranges as well as several city's area ranges.

[3] 10 grainy images of most of the categories of the 2D Hela databank of the US National Institute of Aging were thresholded to various high levels as to obtain 88 files of irregular and separated points.

[4] Cover dataset from the UCI Machine Learning Datasets Repository. 16 files were obtained by selecting the different cover types (*extreme density*).

[5] High resolution (*down to 10-metres accuracy in some areas*) hydrological network and coastal map of North America courtesy of Environment Canada.

[6] 24 clustering datasets of the Speech and Image Processing Unit at the University of Eastern Finland

[7] polygons defining administrative zones, graciously provided by the General Direction of Environment & Land-use Development, Strategic Division, Megève City Hall, France.

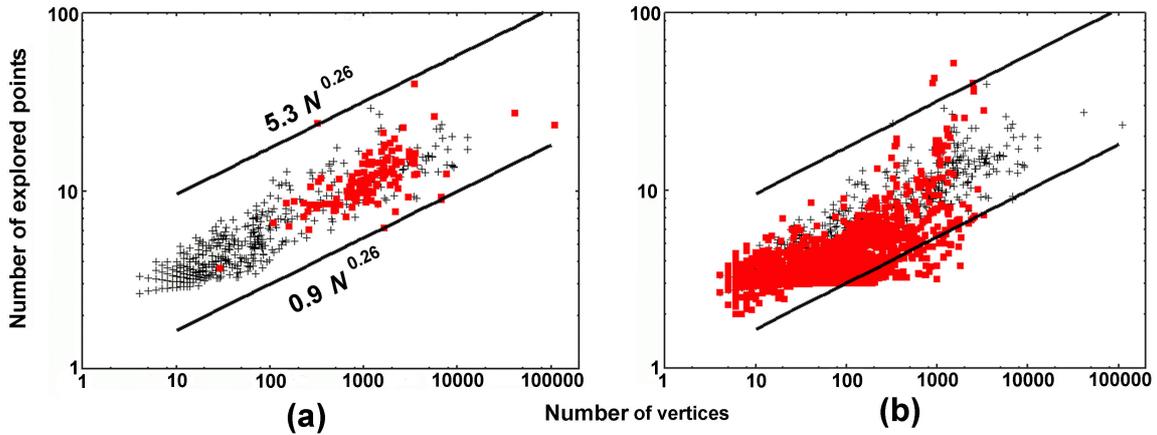

**Figure 5.** Experimental evaluation of the fractional-power value in a scan-line approach

*The plot displays the average number of explored points per segment (i.e. the average number of vertices whose coordinates lie in between those of 2 consecutive hull's vertices in the sorted buffer) versus the number of points in the polygon.*

*(a) Polygons obtained from raw datasets. Black crosses are for lightning data while red dots are for all other sources (the 34 worst-cases excluded). Least-squares constant is 2.2 with a correlation factor of 92.4%.*
*(b) Polygons obtained directly from a GIS. Black crosses are for all polygons obtained from raw datasets while red dots are the GIS-related polygons*

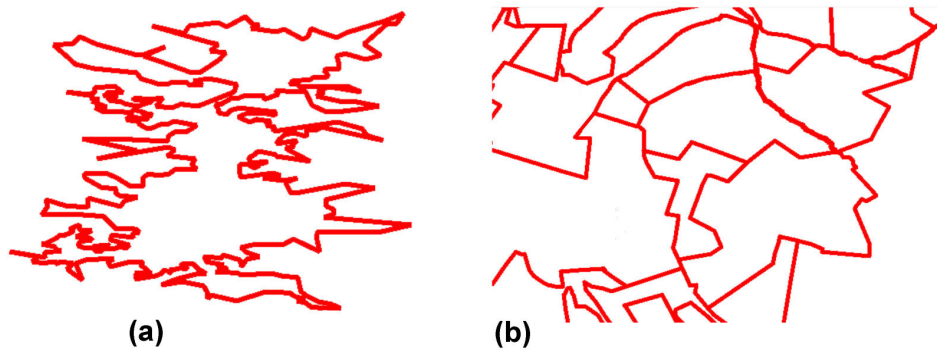

**Figure 6.** The difference in polygons between raw datasets and GIS

*(a) Polygon originating from raw datasets (lightning)*
*(b) A series of polygons originating from a GIS*

### 2.6 Influence of backtracking

Although the number of self-crossings per polygon is highly dependent upon the source one might try to evaluate the influence of backtracking. As mentioned in Paragraph 2.2 if some conditions are met the algorithm backtracks more than the usual −1. In order to estimate what the total impact these backtracking might have on the global process one can study the percentage of more-than-usual backtracking, i.e. the real number of vertices which were studied once normal backtracking is removed, versus the number of vertices.

If $N$ is the number of vertices in the polygon, $N_{\text{crossings}}$ is the number of crossings which were corrected and $N_{\text{real}}$ is the real number of vertices which were studied, the number of above-usual study points is: $N_{\text{supp}} = N_{\text{real}} - N - N_{\text{crossings}}$ . Figure 7 displays the percentage represented by $N_{\text{supp}}$ versus $N$. First one may note that only 1.85% of the whole sample of 15,215 polygons exhibit additional exploration due to unusual backtracking so it should be considered as a marginal and even negligible effect. Then in this small subsample the average value is around 2.5% of the total number of vertices and only one reaches 23%, which is not very significative given the very low number of vertices involved (*it represents only 3 points*).

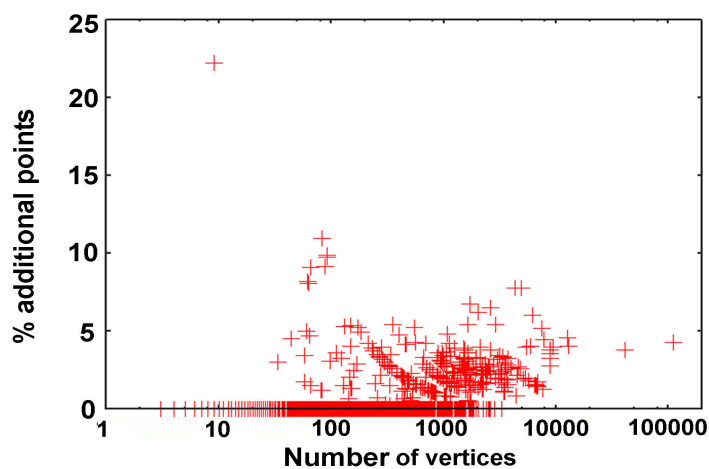

**Figure 7.** Percentage of additional studied points *vs* number of vertices

So it would be safe to conclude that an upper bound for the running time of this algorithm in the average situation is O( $(1.05\ N + k)\ N^{0.26}$ ), if $k$ denotes the number of self-crossings which were corrected: first as above-mentioned the backtracking is marginal and 5% is really an upper bound of the average case; secondly as shown by Figure 5b the exponent might be much lower in average depending upon the origin of the datasets; and finally when correcting $k$ intersections the algorithm might have explored much less than the given average value during the detection phase as it stops as soon as the intersection is found.

### 2.7 Influence of the data structure

When correcting a self-intersection in a polygon some re-ordering of the vertices is present. The data structure used to represent the points might have an impact on this process.

If the input points are represented as an array correcting the intersection will lead to a re-ordering and re-numbering of the vertices. Then if the algorithm is based on a sorted list of these points, once the correction is made on the real points the sorted array will need an update whether the points are referred to by their addresses or by their indexes in the array. In order to do that one has to check all indexes within the range defined by the modified section and update only the relevant ones.

If the input points are represented as a chained list of points on the other hand correcting an intersection will only consist in re-ordering the next and previous pointers of the involved range of points. The points themselves will be unchanged and so will the sorted list. This will in consequence be faster. It will however use 3 times more memory.

*2.8 Influence of the sorting direction*

While most authors use a vertical sorting direction some, as de Berg & al.[5], use an horizontal one. Although not having much of a direct impact on the usual tree-based methods it might have one on the scan-line approach, as one will have to go through a portion of the total number of points for each segment. In this study it was found that in average the difference between using the major coordinate or a fixed one amounted to $N^{0.07}$ in the average case. Thus although not modifying the magnitude of the factor it would be nevertherless best to sort the points along their major coordinate rather than choose one direction for all datatsets. Eventually using the data's major axis could be done rather than using the major coordinate in order to avoid all worst-cases, get all cases around the average and take into account unbalanced distributions for instance. It will however involve heavier computations to compute projections on this axis while exploring the array.

In any case, even if one uses the usual methods for solving this problem, sorting the points along the major direction of the data (*or along their major axis*) might be very useful as one might gain a lot on the accuracy of the intersection tests by increasing the intervals between the values thus avoiding most if not all of the degeneracies.

## 3. Differences with the brute-force algorithm

Whatever the above-mentioned method used for detecting or correcting the intersections, if applied to convert a complex polygon into a simple one an often-overlooked problem should be noted and is shown in Figure 8.

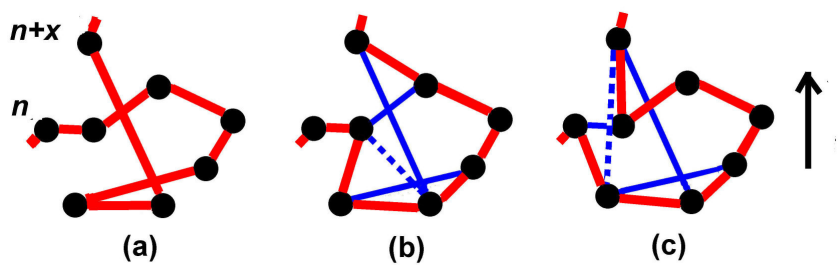

**Figure 8.** Differences on self-intersecting corrections.

*(a) A self-intersecting section of a polygon*
*(b) Correction using the brute-force algorithm*
*(c) Correction using a basic sweep-line algorithm*

*Dashed lines represent the intermediate segments after the first correction.*

As the points are sorted through one of their coordinate (*the arrow in Figure 8 above*) the picking order differs from the brute-force algorithm, and so do the end results. This is quite unusual, and even quite unique, for obtaining a different output whether an algorithm is optimized or not is not a usual trait in algorithmics or computational geometry.

There are three main efficient ways of computing and correcting self-intersections in a polygon using brute-force. They all only check upper indexes but they differ on the handling of what happens when an intersection is corrected: the first one simply backtracks one position then iterates once it has reached the end of the polygon to check for newly created intersections; the second one checks between the two lower indexes of the involved segments for lower intersections; the third one is more logical and more efficient: only the two new formed segments can produce intersections with lower indexes. So one has to check for each of the new segments for a lower intersection, the check for the one having the lowest index being recursive. Here the first method will not be looked at as it has the worst efficiency.

In order to address the problem of the differences induced by the different picking order between optimised methods and the brute-force algorithm one would have to be able to answer the query "*does segment (i) intersects segment (j) ?*" in a way which does not depend upon previous processing. Although some of the methods cited in this paper are able to answer that query in optimal time they will need the tree in order to give the answer. However in general the usual algorithms can answer a query like "*does segment (i) intersects another segment ?*" and, if the answer is positive, *output* the segment's number, with no control over the segment number. Then if that query arose from other parts of the software keeping the tree in memory for this particular object while other computations have taken place might prove quite a burden.

The scan-line approach described in the previous chapter is a good avenue to research as it only needs the sorting of the input points, which could be more easily shared or re-computed. It could be expected however to use a much higher fraction of the points as one does not have the pre-information given by the sequential processing of the buffer. The main difficulty lies in finding *where* to stop the exploration. Although this is quite impossible to find for a set of disjoint segments some heuristics can be found for polygons as vertices form a closed path.

Figure 9a shows an example of a self-intersecting polygon. Assuming that the vertices were sorted by increasing y-coordinate (*in this case*) it can be seen that the polygon is split into two halves, one higher than the studied segment and one lower. Each of these halves consists in two parts: a left and a right one. It thus could be imagined that some criterion based on this double splitting could be found, which will indicate whether the exploration of potential candidates is complete based on what Figure 9b shows.

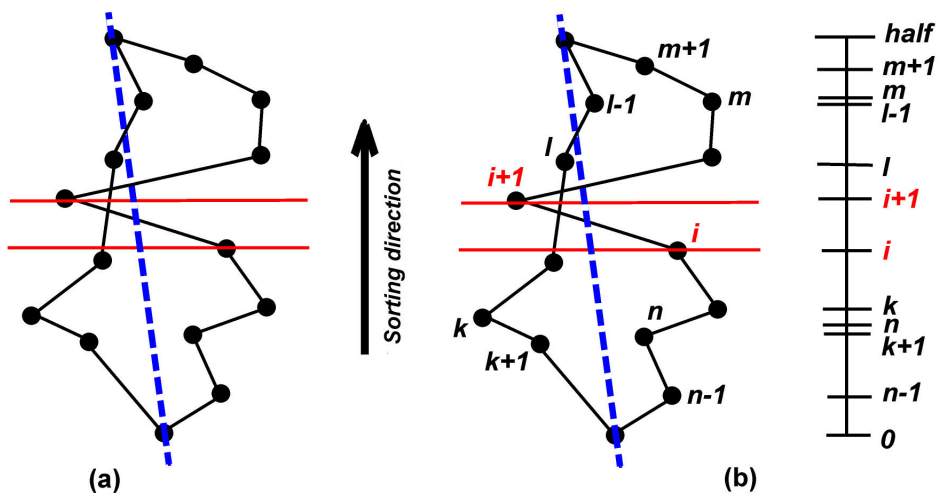

**Figure 9.**. Top/bottom and left/right polygon decomposition

*(a) The red lines cut the polygon vertically in two parts while the dashed blue line, linking the lowest to the highest point according to the sorting direction, cuts the polygon in two other halves, the left one and the right one.*

*(b) Using this double splitting to idenitfy points this shows what the sorted array looks like: n or m are for right-side points while k or l are for left-side points. The +1 or −1 are relative to the studied segment.*

Following Figure 9b the space is divided into five parts: first what lies in between the two segment's extremities then a lower left, a lower right, an upper left and an upper right part. However the difficulty lies in finding *when* a section of the polygon can be said to be "homogeneous" or complete, i.e. the bottom-left part or the up-right part for instance. This is tricky as the "homogeneous" behaviour is contradicted if an index of the other quarter of the same side is found in between two consecutive points. In order to handle this case, every time such a case arises the "homogeneousness" has to be destroyed. This is also the case for segments perpendicular to the sorting direction, leading to the next index lying at the next position in the array. The sample code below demonstrates the tests for the upper-right section:

```
if ( Candidate < Half ) {
    if ( StartUpperRight < 0 ) {                   Part 1
        StartUpperRight = Candidate
        EndUpperRight = Candidate + 1
        NumStartUpperRight = i ;
    }
    else
    if ( Candidate = EndUpperRight ) {             Part 2
        if ( i = (NumStartUpperRight+1) ) {
            StartUpperRight = -1
        }
        else {
            UpperRightComplete = True
        }
    }
    else
    if ( Candidate > StartUpperRight ) {           Part 3
        StartUpperRight = Candidate
        EndUpperRight = Candidate + 1
        NumStartUpperRight = i
        UpperRightComplete = False
    }
    else
    if ( Candidate < StartUpperRight ) {           Part 4
        StartUpperRight = -1
        UpperRightComplete = False
    }
}
```

**Sample code for the completion test of the upper-right section**

*i is the index of the sorted array the algorithm is exploring, pointing to the vertex numbered Candidate. Half corresponds to the vertex number of the last point in the sorted array*

Part 1 deals with either the first point or the next point after a reset. Part 2 deals with an index corresponding to the expected value with a special case for perpendicular segments, for which a reset is done. Part 3 deals with a vertex belonging to the same section but situated above the actual point: limits are then recomputed. Finally Part 4 deals with a point breaking the section's sequence: a reset is also done.

The four corner sections of the space division will be explored through the same algorithm, with only sign changes in the end point computation as well as in the tests. Although the reasoning at the core of the sweep- or scan-line-based methods is purely in one direction, in this case one has to assume that exploration should go in both directions in order not to miss any possibility: because of the lack of prior pre-preprocessing all four sections will have to be complete before ending the exploration

So the algorithm should go as follows:

- First the "scan-line interval" is explored, i.e. all points lying in between the segment's extremities in the sorted array are checked.

- Then points situated below the lowest segment's extremity in the sorted array are checked. Exploration stops when both sides have an homogenous portion lying below.

- Then points situated above the highest segment's extremity in the sorted array are checked. Exploration stops when both sides have an homogenous portion lying above.

It should be noted also that during these checkings one can select whether only segments with higher – or lower - index values are to be taken into account.

Obviously the fractional-power factor will be much higher than the basic scan-line one. An experimental study of this type of method on the above-mentioned datasets indeed leads to an asymptotic value of 0.65 with a 100% similarity rate with brute-force end results of either methods. If the majority of points is in the direction perpendicular to the sorting one the factor reaches 0.8. It is however possible to reach an *average* value of 0.6 by entering the Part 3 reset of the above-mentioned algorithm only if the studied section is not already completed:  it leads to a 99.999% similarity rate if an algorithm based on the third method is used while keeping 100% if the second method is used.

Although not satisfactory for the complete processing of a self-intersecting polygon it could be of use for test purposes or to answer several unrelated queries on particular segments, as it is nervertherless a good speed optimisation, i.e. $O(N^{0.65})$ compared to $O(N)$, allowing for instance to check only just around 1,800 segments instead of 100,000 for the brute-force.

## 4. Conclusion

This paper has presented some thoughts on the memory consumption aspects of the usual optimal methods used to list the intersections of a list of segments or a self-intersecting polygon. A scan-line based algorithm is proposed using only the minimal memory space required by the sorting, the listing running in $O(N^{1.26})$ or the correction in $O((N+k)\ N^{0.26})$ time complexity at most in average, with a high probability of a much lower exponent around 0.16 and even as low as 0.1, and based on a very simple and easy-to-implement code. It could therefore be of great help for inclusion in much larger software for which this computation is only a small part of the whole, and is also well suited for very large datasets or environments for which memory consumption is a major constraint. Then differences between the usual methods and the brute-force algorithm were noted and a work-around algorithm is proposed to obtain similar results which, while being in $O((N+k)\ N^{0.65})$ time complexity at best, could still be used with advantage mainly to speed up isolated queries on a segment's possible intersections.

## Acknowledgments

The author wishes to thank Loic Bartoletti, from the General Direction of Environment & Land-use Development, Strategic Division, Megève City Hall, France, for his invaluable help in providing a large set of self-intersecting polygons for test purposes.

JeanSouviron@hotmail.com